\documentstyle[12pt,epsfig]{article}
\newcommand{\preprint}[1]{\begin{table}[t]  
           \begin{flushright}               
           \begin{large}{#1}\end{large}     
           \end{flushright}                 
           \end{table}}                     

\begin{document}

\preprint{chao-dyn/9809017\\TAUP-2523-98}

\begin{center}
{\large{\bf Instability and Chaos in Non-Linear Wave

Interaction: a simple model.}}

\vspace{50pt}

\bf{I. M. Khalatnikov$^{a,b}$ and M. Kroyter$^{a,}$\footnote{email: mikroyt@post.tau.ac.il}}

\vspace{20pt}

{\it $^a$  School of Physics and Astronomy,

 Raymond and Beverly Sackler Faculty of Exact Sciences,

Tel Aviv University, 69978 Tel Aviv,

Israel}

\vspace{15pt}

{\it $^b$  L. D. Landau Institute for Theoretical Physics,

Russian Academy of Sciences,

2 Kosygin Street, Moscow, 117940,

Russia}

\end{center}

\vspace{20pt}

\begin{abstract}
We analyze stability of a system which contains an harmonic oscillator non-linearly coupled
to its second harmonic, in the presence of a driving force. It is found that there always
exists a critical amplitude of the driving force above which a loss of stability appears.
The dependence of the critical input power on the physical parameters is analyzed.
For a driving force with higher amplitude chaotic behavior is observed.
Generalization to interactions which include higher modes is
discussed.

{\it keywords: Non-Linear Waves, Stability, Chaos.}
\end{abstract}

\newpage

\section{Introduction}

In a series of experiments the motion of the surface of a superfluid liquid in a cylindrical
vessel was studied. This motion was induced by standing waves of second sound propagating
in the bulk of the liquid. Above a critical value of the input power the motion has lost
stability \cite{Olsen,Egolf}.

To account for this loss of stability we analyzed a model that explained this phenomenon
\cite{Kroyter}, and found it to be in a good agreement with the experimental results.
The model is general enough to account for the loss of stability in
other wave systems.

\section{The Model}

The model consists of two non-linearly coupled harmonic oscillators, of which one is
coupled to an external driving force. First we would like to justify the use of two oscillators,
with frequencies close to $\omega$ and $2\omega$, for describing the physics of systems
such as the one above ($\omega$ would be the frequency of the driving force).
We assume that in the linear approximation the free, non-dissipative (classical) theory is given
by the Hamiltonian:

\begin{equation}
H=\sum_{n=1}^\infty \omega_n a^*_n a_n
\end{equation}
where $a_n$ is the (complex) amplitude of the $n^{th}$ mode, and $a^*_n$ is its complex
conjugate. Dissipation and driving force would be added in the following.
The modes are the eigenfunctions of the wave equation with the appropriate Sturm-Liouville
boundary conditions.

We neglect terms higher then cubic from the Hamiltonian, as well as terms which
are far from resonance, and therefore have small coupling constant \cite{Lvov}.
The Hamiltonian turns to:

\begin{equation}
\label {eq:fullH} H=\sum_{n=1}^\infty \omega_n a^*_n a_n+\sum_{\stackrel{k+l-m \simeq 0}
{k,l,m=1}}^\infty (\lambda_{k,l;m} a_k a_l a^*_m+c.c.)
\end{equation}
where c.c. stands for complex conjugate, and $\lambda_{k,l;m}=\lambda_{l,k;m}$.
We would like now to couple an external driving force to one of the modes. We keep in mind
that, in order to describe a physical problem, attenuation should be added as well. The modes
which are not strongly coupled to the excited mode would decay.
Again we assume that, for
describing the onset of instability, a minimal number of modes is needed.
Therefore we
take the excited mode and the mode with frequency which is closest to twice the frequency
of the first one. With the harmonic driving force the Hamiltonian takes the form:

\begin{equation}
H=\omega_d a_d^* a_d+\omega_{2d} a^*_{2d} a_{2d}+(\lambda a_d^2 a^*_{2d}+c.c.)+
(f e^{i \omega t}a_d^* + c.c.) \label{eq:partH}
\end{equation}
where $\omega$ is the frequency of the driving force,
which should be close to $\omega_d$ in order to resonate it.

We use:

\begin{equation}
i \dot a_d={\partial H \over \partial a_d^*}
\end{equation}

which is Hamilton's equations in the amplitude formalism \cite{Lvov},
to derive the equations of motion:

\begin{eqnarray}
i \dot a_d & = & \omega_d a_d+2\lambda^* a_d^* a_{2d}+f e^{i \omega t} \\
i \dot a_{2d} & = & \omega_{2d} a_{2d}+\lambda a_d^2
\end{eqnarray}

The equations are invarant under the transformation:

\begin{eqnarray}
\label{eq:inv1}
\begin{array}{rcl}
a_d & \rightarrow &a_d e^{i (\phi+\theta)} \\
a_{2d} & \rightarrow &a_{2d} e^{i (\phi-\theta)} \\
\lambda & \rightarrow &\lambda e^{i (-\phi-3\theta)} \\
f & \rightarrow &f e^{i (\phi+\theta)}
\end{array}
\end{eqnarray}

It is therefore possible to eliminate two independent phases from the equations,
so we can choose $\lambda$ and $f$ to be real.

We add now dissipative terms to the equation in the usual maner \cite{Lvov}. The equations
now become:

\begin{eqnarray}
i \dot a_d & = & (\omega_d-i \gamma_d) a_d+2\lambda a_d^* a_{2d}+f e^{i \omega t} \\
i \dot a_{2d} & = & (\omega_{2d}-i \gamma_{2d}) a_{2d}+\lambda a_d^2
\end{eqnarray}
where $\gamma$ are the dissipation constants.

The final stage before analyzing the equations is to introduce the ``slow variables'' to
eliminate the time dependence. Under the transformation:

\begin{eqnarray}
\nonumber a_d & \rightarrow &a_d e^{-i \omega t} \\
\nonumber a_{2d} & \rightarrow &a_{2d} e^{-2i \omega t}
\end{eqnarray}

the equations get the form:

\begin{eqnarray}
\label{eq:1} i \dot a_d & = & (\Delta_d-i \gamma_d) a_d+2\lambda a_d^* a_{2d}+f \\
\label{eq:2} i \dot a_{2d} & = & (\Delta_{2d}-i \gamma_{2d}) a_{2d}+\lambda a_d^2
\end{eqnarray}

where $\Delta_d \equiv \omega_d-\omega$ and $\Delta_{2d} \equiv \omega_{2d}-2\omega$
are the frequency detunings with respect to the driving force.

We note that in deriving the Hamiltonian (\ref{eq:partH}) we have neglected
one non-linear term which is of the same order with respect to $a_d$ as the one
that we have kept, namely $\kappa a_d^2 a_d^{*2}$. We would like to justify it.
For systems in which $\kappa << {\lambda^2 \over \omega}$ this term is small,
but it turns out that even for $\kappa$ much larger the importance of this
term is not crucial. Note, that $\omega$ is absent in the equations.
From dimensional considerations $\kappa$ can appear at the equations only
as $\kappa \Delta$, $\kappa \gamma$, this is the term that
has to be of the order of $\lambda^2$.
Hereafter we analyze three aspects of the model: stationary solutions,
stability, numerical calculations.
For the stationary solutions it is easy to verify that the effect of
$\kappa$ is merely to renormalize $\Delta_{2d},\gamma_{2d}$. This is
the well known effect of shifting the resonance \cite{Landau}.
We have seen that $\kappa$ is not of a big importance,
even for $\kappa>{\lambda^2 \over \Delta_{2d}},{\lambda^2 \over \gamma_{2d}}$,
in the stability analysis as well as in our numerical calculations.
We will not include this term in what follows.

Although the model we use is a very simplified one, it still contains five
parameters in addition to the driving force amplitude $f$. Not all
the parameters are important.
The amplitude $f$ of the driving force is an effective expression which
in fact is a function of $\Delta_d$, moreover, the driving force couples
to all other modes as well, and we may neglect all other couplings only
when the one that we are left with is the dominant one.
For this to be the case we must have $\omega \simeq \omega_d$, that is,
$\Delta_d$ is small compared to all other parameters with dimensions of
frequency. The value of $\Delta_{2d}$ will be dictated by geometry.
Both, our analytical, as well as numerical results depend on this
assumption. In most physical systems there is a relation between
$\gamma_d$ and $\gamma_{2d}$. We shall assume that this two parameters
are of the same order of magnitude.

\section {Stationary solutions}

We begin our analysis by finding the fixed points of the equations, i.e., solving the
equations:

\begin{eqnarray}
(\Delta_d-i \gamma_d) a_d+2\lambda a_d^* a_{2d}+f &=& 0 \\
(\Delta_{2d}-i \gamma_{2d}) a_{2d}+\lambda a_d^2 &=& 0
\end{eqnarray}
We eliminate $a_{2d}$ from the second equation, and substitute in the first one to get:

\begin{equation}
(\Delta_d-i \gamma_d)(\Delta_{2d}-i \gamma_{2d}) a_d-2\lambda^2 a_d |a_d|^2 =
-(\Delta_{2d}-i \gamma_{2d})f
\end{equation}
The equation for $\zeta \equiv {2\lambda^2 \over |\gamma_d \gamma_{2d}-
\Delta_d\Delta_{2d}|} |a_d|^2$
turns now to:

\begin{equation}
\label{eq:quad} ((\zeta \pm 1)^2+\beta)\zeta=h
\end{equation}
where:

\begin{equation}
h \equiv {2(\gamma_{2d}^2+\Delta_{2d}^2) \lambda^2 \over
 |\gamma_d \gamma_{2d}-\Delta_d \Delta_{2d}|^3}f^2
\end{equation}
is the scaled force, and

\begin{equation}
\label{eq:beta} \beta \equiv ({\Delta_d \gamma_{2d}+
        \Delta_{2d} \gamma_d \over \gamma_d \gamma_{2d} - \Delta_d \Delta_{2d} })^2
\end{equation}

The sign in equation (\ref{eq:quad}) coincides with the sign of
$\gamma_d \gamma_{2d} - \Delta_d \Delta_{2d}$.

This equation has either one or three solutions. For a given value of $h$ the equation
would have three solutions if and only if:

\begin{eqnarray}
\label{eq:gg} && \gamma_d \gamma_{2d} - \Delta_d \Delta_{2d}<0 \\
\label{eq:beta1}&& 0 \leq \beta < {1\over3} \\
\label{eq:ns}&& {2 \over 27}[1+9 \beta-(1-3\beta)^{3 \over 2}] \leq h \leq 
 {2 \over 27}[1+9 \beta+(1-3\beta)^{3 \over 2}]
\end{eqnarray}

In the following, it would be illustrated that, when three solutions are
present, the middle one is non-stable, as may be expected.

We note that the situation of three solutions is, in a sense, non-physical.
$\gamma_d$, $\gamma_{2d}$ are positive, we can therefore use
(\ref{eq:beta})-(\ref{eq:beta1}) to deduce:

\begin{equation}
({\gamma_d \over \Delta_d})^2<({\gamma_{2d} \over \Delta_{2d}} +
         {\gamma_d \over \Delta_d})^2<{1\over3}
\end{equation}

But this suggests that $\gamma_d<\Delta_d$, which contradicts our assumptions.
In this region of parameters our model is not adequate.

\section {Stability}

To check whether the stationary solutions are stable we linearize the equations around these
solutions, and check whether small perturbations grow or decay.
To simplify the calculations, we recall the symmetry (\ref{eq:inv1}) and use it with
$\phi+3\theta=0$ to redefine the stationary value of the first mode, $a^{(0)}_d$, to be real,
without altering $\lambda$. The change of $f$ is not important since $f$ will be absent from
the linearized equations. We substitute in the linearized equations:

\begin{equation}
\label {eq:lin} a_{2d}^{(0)}= -{\lambda {a_d^{(0)}}^2 \over \Delta_{2d}-i \gamma_{2d}}
\end{equation}

The stability is now determined
by $a_d^{(0)}$. Also, to simplify the notations, we will use $a_d$, $a_{2d}$ rather then
$\delta a_d$, $\delta a_{2d}$ for the deviations from the stationary solution.

The linearized equations are:
\begin{eqnarray}
i \dot a_d & = & (\Delta_d-i \gamma_d)a_d + 2\lambda(a^{(0)}_d a_{2d}-
        {\lambda {a^{(0)}_d}^2 \over \Delta_{2d}-i \gamma_{2d}} a_d^*) \\
i \dot a_{2d} & = & (\Delta_{2d}-i \gamma_{2d}) a_{2d}+ 2\lambda a^{(0)}_d a_d
\end{eqnarray}

multiplying by $-i$ and separating to real and imaginary parts, we get the 
differential equation:

\begin{equation}
\begin{array}{l}
\frac{\hbox{\Large \it d}}{\hbox{\Large \it dt}}
\pmatrix{Re(a_d) \cr Im(a_d) \cr Re(a_{2d}) \cr Im(a_{2d})}  =  \\
\pmatrix{
-\gamma_d-p \gamma_{2d} &
\Delta_d+p \Delta_{2d} &
0 & 2a_d^{(0)} \lambda \cr
-\Delta_d+p \Delta_{2d}&
-\gamma_d+p \gamma_{2d}&
-2a_d^{(0)} \lambda & 0 \cr
0 & 2a_d^{(0)} \lambda & -\gamma_{2d} & \Delta_{2d} \cr
-2a_d^{(0)} \lambda & 0 & -\Delta_{2d} & -\gamma_{2d}   }
\pmatrix{Re(a_d) \cr Im(a_d) \cr Re(a_{2d}) \cr Im(a_{2d})}
\end{array}
\end{equation}
where $p={2{a_d^{(0)}}^2 \lambda^2 \over \gamma_{2d}^2+\Delta_{2d}^2}$.

To ensure stability we shall require that the real part of all the eigen-values of 
this matrix is negative. We find the coefficients of the characteristic polynomial
$u^4+au^3+bu^2+cu+d$, to be:

\begin{eqnarray}
\label{eq:da} a& =& 2(\gamma_d+\gamma_{2d}) \\
\label{eq:db} b& =& -{4 \lambda^4 \over \gamma_{2d}^2+\Delta_{2d}^2} {a_d^{(0)}}^4 + 
        8\lambda^2 {a_d^{(0)}}^2+(\gamma_d^2+\Delta_d^2+\gamma_{2d}^2+\Delta_{2d}^2+
                4\gamma_d \gamma_{2d}) \\
\label{eq:dc} c& =& -{8 \lambda^4 \gamma_{2d} \over \gamma_{2d}^2+\Delta_{2d}^2} {a_d^{(0)}}^4 + 
        8\lambda^2 (\gamma_d+\gamma_{2d}){a_d^{(0)}}^2+ \cr & &+
                2[(\gamma_d^2+\Delta_d^2)\gamma_{2d}+(\gamma_{2d}^2+\Delta_{2d}^2)\gamma_d] \\
\label{eq:dd} d& =& 12\lambda^4 {a_d^{(0)}}^4 +
        8\lambda^2 (\gamma_d \gamma_{2d}-\Delta_d\Delta_{2d}) {a_d^{(0)}}^2 +
        (\gamma_d^2+\Delta_d^2)(\gamma_{2d}^2+\Delta_{2d}^2)
\end{eqnarray}

To ensure that all the roots of this polynomial have negative real part we use the
Hurwith-Routh criterion \cite{Guillemin}\cite{Uspensky}:

\begin{eqnarray}
\label {eq:a}   a&>&0\\
\label {eq:b}   b&>&0\\
\label {eq:d}   d&>&0\\
\label {eq:cc}  abc&>&c^2+a^2d
\end{eqnarray}

The condition (\ref {eq:a}) is trivial for a physical problem.
The condition (\ref {eq:b}) is a quadratic equation with respect to ${a_d^{(0)}}^2$, and is
easily solved to give:

\begin{equation}
\label {eq:bsol} {a_d^{(0)}}^2<{\gamma_{2d}^2+\Delta_{2d}^2 \over 4 \lambda^2}(4+\sqrt{5+
        {\gamma_d^2+\Delta_d^2+4\gamma_d \gamma_{2d} \over \gamma_{2d}^2+\Delta_{2d}^2}})
\end{equation}

The third condition, (\ref {eq:d}), is again a quadratic equation with 
respect to ${a_d^{(0)}}^2$, but with positive, rather then negative coefficient of $a_d^4$. 

It is easily seen that for a negative $d$ to occur at the physical region: ${a_d^{(0)}}^2>0$,
we need to have:

\begin{equation}
\gamma_d \gamma_{2d}>\Delta_d \Delta_{2d}
\end{equation}

When this condition is fulfilled, an unstable region appears when:

\begin{equation}
\beta<{1\over3}
\end{equation}

Direct solution of the quadratic equation then shows that the central region of solutions
coincides exactly with this unstable region (\ref{eq:ns}).
As mentioned above, this region is not physically important.

We combine (\ref{eq:da})-(\ref{eq:dd}) and (\ref{eq:cc}), and
define:

\begin{equation}
z=\lambda^2 {a_d^{(0)}}^2
\end{equation}

to get the last inequality:

\begin{equation}
\label{eq:lastineq} a_0 z^4+a_1 z^3+a_2 z^2+a_3 z+a_4>0
\end{equation}

where:

\begin{eqnarray}
&&a_0={64\gamma_d \gamma_{2d} \over (\gamma_{2d}^2+\Delta_{2d}^2)^2} \\
&&a_1=-{64(\gamma_d+\gamma_{2d})^2 \over \gamma_{2d}^2+\Delta_{2d}^2} \\
&&a_2={32\gamma_d \gamma_{2d} \over \gamma_{2d}^2+\Delta_{2d}^2}
        (\Delta_{2d}^2-\Delta_d^2-(\gamma_d+\gamma_{2d})^2) \\
&&a_3=16(\gamma_d+\gamma_{2d})^2[(\gamma_d+\gamma_{2d})^2+(\Delta_d+\Delta_{2d})^2)] \\
&&a_4=4 \gamma_d \gamma_{2d} [(\gamma_d+\gamma_{2d})^2+(\Delta_d+\Delta_{2d})^2]
        [(\gamma_d+\gamma_{2d})^2+(\Delta_d-\Delta_{2d})^2)]
\end{eqnarray}

It is seen that for all parameter values there exists an open neighborhood of zero
in which the stationary solution is stable. It is very tedious to solve the inequality for
the general case. We solve it for two special cases,
the one-dimensional geometry, and the cylindrical wave with a large Q-factor.
Both with reflecting boundary conditions.

We remind the assumption: $\Delta_d<<\gamma_d$.
It is natural to assume that $\gamma_d$ and $\gamma_{2d}$
are of the same order of magnitude. In a wide class of cases $\gamma \propto \omega^2$,
and therefore:
\begin{equation}
\label{eq:g4g}\gamma_{2d}\simeq 4\gamma_d
\end{equation}
We shall consider this case for both geometries. 
The value of $\Delta_{2d}$ is dictated by geometry. 

For the one dimensional geometry the $d^{th}$ mode is $\cos(d{\pi x \over L})$,
where $L$ is the length of the vessel. This dependence gives:
\begin{equation}
\begin{array}{ll}
\Delta_{2d} & =\omega_{2d}-2\omega=2\Delta_d-(2\omega_d-\omega_{2d})=2\Delta_d-
c(2k_d-k_{2d})=\\
&=2\Delta_d-c(2{\pi d \over L}-{\pi 2d \over L})=2\Delta_d
\end{array}
\end{equation}

We therefore, have for the one-dimensional case:
\begin{equation}
\label{eq:flat}\Delta_d,\Delta_{2d}<<\gamma_d,\gamma_{2d}
\end{equation}

We define:

\begin{eqnarray}
x={z \over \gamma_d^2} \\
s={\gamma_{2d} \over \gamma_d}
\end{eqnarray}

We use (\ref{eq:flat}) to get the inequality:

\begin{equation}
x^4-s(1+s)^2 x^3 - {1 \over 2}s^2(1+s)^2 x^2 +{1\over 4}s^3(1+s)^4  x +
{1 \over 16} s^4(1+s)^4   >0
\end{equation}

The solution of this inequality combined with (\ref{eq:bsol})
gives the final result:

\begin{equation}
x<{1\over2}(s+s^2)
\end{equation}
from which one easily finds an expression for the critical input power:

\begin{equation}
f_c={3\gamma_d+\gamma_{2d} \over \lambda} \sqrt{\gamma_{2d}
{\gamma_d+\gamma_{2d} \over2}}
\end{equation}
or using (\ref{eq:g4g}):
\begin{equation}
f_c\simeq{22 \over \lambda} \gamma_d^2
\end{equation}
we now substitute $\gamma_d=\alpha \omega^2$ to get:
\begin{equation}
f_c\simeq{22\alpha^2 \over \lambda} \omega^4
\end{equation}

A full description of the loss of stability for the specific problem can be obtained
if we take into account the dependence of $\alpha$ and $\lambda$ on the relevant physical
parameters, i.e. temperature.

For a cylindrical vessel of radius $R$ the modes are given by $J_n(kr)\cos(n \theta)$,
where $J_n$ is the $n^{th}$ Bessel function,
and $k= {\omega \over c}$ where $c$ is the wave velocity. The boundary conditions force the
relation $k_{n,m}R=\chi_{n,m}$ where $\chi_{n,m}$ is the $m^{th}$ zero of $J_n'(\chi)$.
For simplicity we shall consider here only the $J_0$ modes.

The value of $\Delta_{2d}$ is dictated by the
Bessel asymptotics:
\begin{equation}
\label {eq:bessel} \chi_m \equiv \chi_{0,m} \simeq n\pi + {\pi \over 4}
\end{equation}
using which we get:
\begin{equation}
\begin{array}{ll}
\Delta_{2d} & =\omega_{2d}-2\omega=2\Delta_d-(2\omega_d-\omega_{2d})=2\Delta_d-
c(2k_d-k_{2d}) \simeq \\
& \simeq 2\Delta_d-{c \over R}(2\chi_d - \chi_2d)=2\Delta_d-{c \over R}
(2(d\pi+{\pi \over 4})-(2d\pi+{\pi \over 4}))=\\
& =2\Delta_d-{\pi c \over 4R} \simeq 2\Delta_d- {\omega_d \over 4d+1}
\end{array}
\end{equation}

Since $\omega_d= 2Q \gamma_d$, the higher is $Q$, the higher are the values of $d$ for which 
the inequality
\begin{equation}
\Delta_{2d}>>\gamma_d
\end{equation}
holds.

We solve now equation (\ref{eq:lastineq}) for the case:
\begin{equation}
\Delta_d<<\gamma_d,\gamma_{2d}<<\Delta_{2d}
\end{equation}

We define $s$ as before, but now:
\begin{equation}
x={z \over \Delta_{2d}^2} 
\end{equation}
and get the equation:
\begin{equation}
x^4-{(1+s)^2 \over s} x^3 + {1 \over 2} x^2 +{(1+s)^2 \over 4s} x +{1 \over 16} >0
\end{equation}

When this condition is combined with (\ref{eq:bsol}) we get:
\begin{equation}
x<{1\over4}(v+\sqrt{2uv})
\end{equation}
where $u={(1+s)^2 \over s}$, and $v=u-\sqrt{u^2-4}$. We use (\ref{eq:g4g}) to
get: $x<0.59$. For other values of $s$ there are only small changes in the result. In all
cases the critical value is in the range:
$0.5<x_0<0.65$. The maximal value is attained at $s=1$, and the minima are at $x=0$,
$x \rightarrow \infty$ (note that $x_0(s)=x_0({1 \over s})$).
The critical input power $f_c$ may be calculated now:
\begin{equation}
f_c={2 \Delta_{2d}^2 x_0^{1.5} \over \lambda} \simeq 0.56 {c^2 \over \lambda R^2}
\end{equation}

A full description of the loss of stability in this geometry can be obtained
if we take into account the dependence of $\lambda$ and $c$ on the relevant physical parameters.

\section {Beyond - Numerical calculations.}

Some questions arise. Does the system always reach the stationary solution in the stable region? 
What happens above the stable region? In what way would the theory be modified if we include
the full Hamiltonian (\ref{eq:fullH})?

We solved the equation numerically with parameters suitable to describe
the cylindrical geometry:
\begin{equation}
\begin{array}{ll}
\Delta_d=0 &\Delta_{2d}=1500 \\
\gamma_d=30 &\gamma_{2d}=120 \\
\lambda=5400
\end{array}
\end{equation}

with initial conditions:
\begin{equation}
\begin{array}{ll}
a_d(t=0)=0 & a_{2d}(t=0)=0
\end{array}
\end{equation}
The results for
other values of parameters may be very similar due to the scaling properties
of the equation discussed above.

\begin{figure}[hbtp]
\begin{center}
\mbox{\epsfig{figure=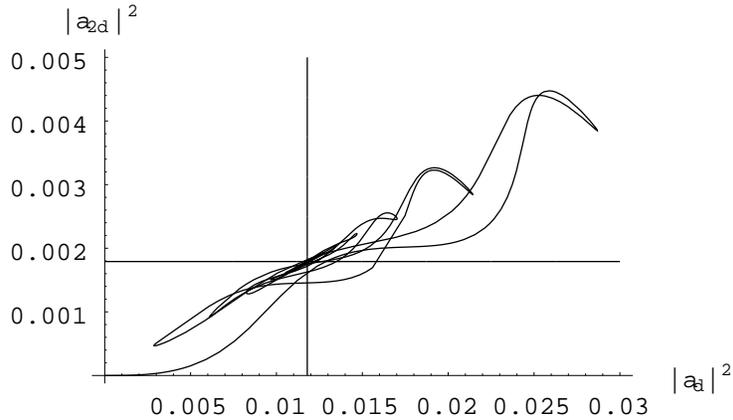,width=10cm}}
\end{center}
\caption{The system approaches the fixed point for $f=50$, the position of
the fixed point is indicated.}
\label{fig:point}
\end{figure}

\begin{figure}[hbtp]
\begin{center}
\mbox{\epsfig{figure=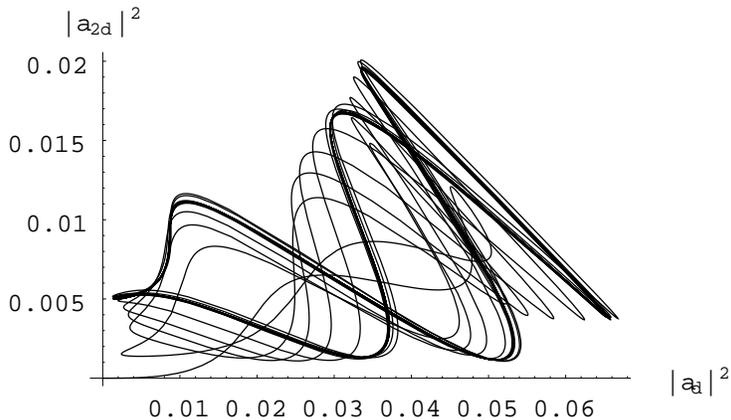,width=10cm}}
\end{center}
\caption{The system approaches a limit cycle for $f=100$ which is below the
critical value.}
\label{fig:cycle}
\end{figure}

For small enough values of $f$ the system reaches the stationary solution
after some travelling in phase space (Fig. \ref{fig:point}).
For $f \simeq 0.3f_c$ with the initial conditions above, the system escapes the basin of
attraction of the fixed point, and rather approaches a limit cycle
(Fig. \ref{fig:cycle}).
The basin of attraction of the stationary solution shrinks
to zero as the instability is approached. This limit cycle is not unique.
By choosing various initial conditions other limit cycles can be approached.
In the higher $f$ regime the behavior is harder to determine.

It is easy to prove that the motion of the system is bounded in its phase space,
and that the volume in phase space decays exponentially with decay factor
$2(\gamma_d+\gamma_{2d})$.

A necessary condition for chaos to evolve is that the system will be
unstable locally.
Our analysis shows that the phase of $a_d$, $a_{2d}$ is irrelevant to this question.
Given the value of the parameters, the potentially chaotic regions are
defined in the $(|a_d|^2,|a_{2d}|^2)$ plane.
Our calculations show that the region
\begin{equation}
\label{eq:chaos} |a_{2d}|^2\lambda^2>>\gamma^2, \Delta^2
\end{equation}
is always locally unstable.
The numerical calculations show that when $f$ is increased
the system enters this region, bifurcations appear, as in the usual root towards 
chaos. For large enough $f$ chaos will evolve.

\begin{figure}[hbtp]
\begin{center}
\mbox{\epsfig{figure=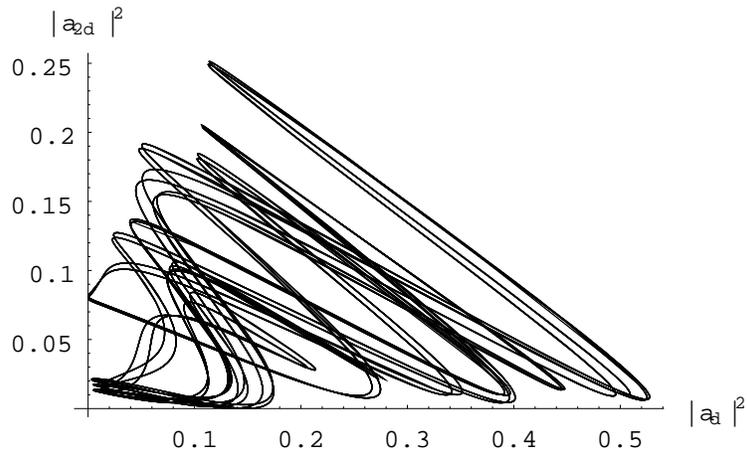,width=10cm}}
\end{center}
\caption{For $f=500$. One of the limit cycles which bifurcates towards chaos.}
\label{fig:chaos500}
\end{figure}

In (Fig. \ref{fig:chaos500}) we see the bifurcations for $f=500$.

\begin{figure}[hbtp]
\begin{center}
\mbox{\epsfig{figure=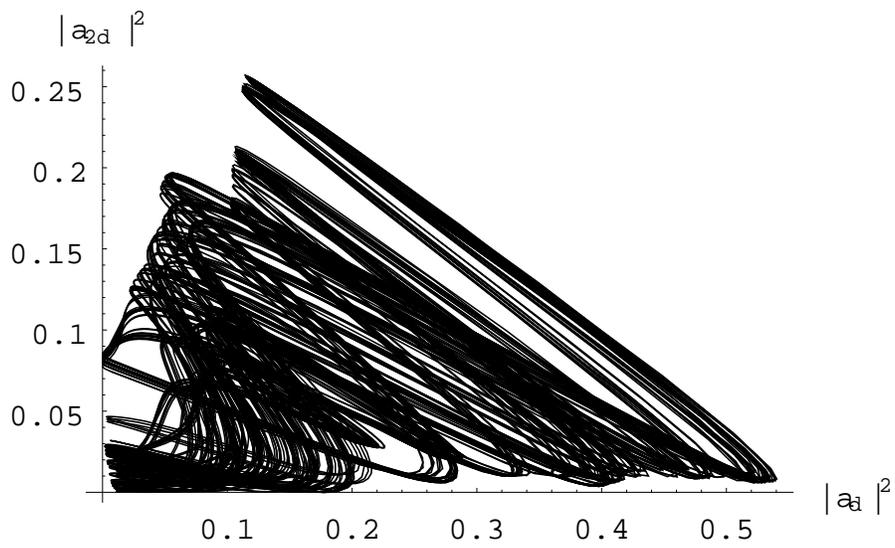,width=12cm}}
\end{center}
\caption{For $f=507$ the system is chaotic.}
\label{fig:chaos507}
\end{figure}

Chaos evolves for $f\simeq 506$ as we see in (Fig. \ref{fig:chaos507}).

When more modes are added to the system the behavior changes. The projection of, say,
the 3-mode system on the $(4dim)$ phase space of 2 modes gives, in general, trajectories which
are very different from the original ones. Yet, we argue that the main conclusion
does not change. Indeed, if we examine the original set of equations (\ref{eq:1}),
(\ref{eq:2}) we note that the transformation:
\begin{equation}
\begin{array}{l}
a_d \rightarrow \alpha a_d \\
a_{2d} \rightarrow \alpha a_{2d} \\
f \rightarrow \alpha f \\
\lambda  \rightarrow {1 \over \alpha} \lambda \\
\end{array}
\end{equation}
which is a generalization of (\ref{eq:inv1}), leaves the equations invariant. We could deduce
from here that $f_c \propto {1 \over \lambda}$. From dimensional considerations $f$ should be
proportional to $\gamma^2,\Delta^2$. It is
seen that for the one dimensional case the leading behavior is: $f_c \propto \gamma^2$,
while for the large Q-factor case $f_c \propto \Delta_{2d}^2$.
All our calculations were in fact needed just to illustrate that there is only one transition
from stability to instability (i.e. no instable windows),
to validate the assumption that the largest constant with
frequency dimensions is not absent from the expression for $f_c$, and to calculate $x_0$.
When we add new modes, new constants are added to the system. Since from (\ref{eq:bessel})
we get that for all $j$ $\Delta_j \propto {c \over R}$, these constants do not cause a problem.
The same is true for the one-dimensional case.
As for the new $\lambda's$, if they scale in some way, e.g. if 
\begin{equation}
\lambda_{k,l;m}=f(T,R,...)h({l \over k}, {m \over k})k^u
\end{equation}
where $f(T,R,...)$ is any function of all physical parameters, but
the wave length, $h({l \over k}, {m \over k})$ are constants,
and $u$ is an exponent, then the symmetry still holds and then, given that
the general picture remains the same, all that we need to change is the value of $x_0$.
This necessary modification of $x_0$, plus the shrinking of basin of attraction, which
effectively lowers $x_0$, suggests that this part of our calculations is not reliable.
Yet, the dependence of the critical input power on all physical parameters remains the same even
for the full Hamiltonian (\ref{eq:fullH}). There are values of $u$,
$h({l \over k}, {m \over k})$
for which other predictions, such as the distribution of the chaotic regions of the
2-mode system would not be dramatically changed as well.
More extensive investigation of this system is, therefore, highly desirable.
\section*{Acknowledgments}
The authors would like to express their gratitude to Prof. Boris Chirikov for his kind interest in this work, and for his many important remarks.
One of the authors (I. M. K.) would like to thank Prof. Valery Pokrovsky,
with whom the
problem analyzed in this work was first formulated.
We wish to thank Prof. Naum Meiman for useful discussions.

\end{document}